# Nonlinear dynamics of microtubules – A new model

Slobodan Zdravković[a,*], Miljko V. Satarić,[b] Slobodan Zeković[a]

[a] *Institut za nuklearne nauke Vinča, Univerzitet u Beogradu, Poštanski fah 522, 11001 Beograd, Serbia*
[b] *Fakultet tehničkih nauka, Univerzitet u Novom Sadu, 21000 Novi Sad, Serbia*

A B S T R A C T

In the present paper we describe a model of nonlinear dynamics of microtubules (MT) assuming a single longitudinal degree of freedom per tubulin dimer. This is a longitudinal displacement of a dimer at a certain position with respect to the neighbouring one. A nonlinear partial differential equation, describing dimer`s dynamics within MT, is solved both analytically and numerically. It is shown that such nonlinear model can lead to existence of kink solitons moving along the MTs. Internal electrical field strength is calculated using two procedures and a perfect agreement between the results is demonstrated. This enabled estimation of total energy, kink velocity and kink width. To simplify the calculation of the total energy we proved a useful theorem.

[*]Corresponding author.
 *E-mail address:* szdjidji@vinca.sr

# 1. Introduction

Microtubules are major cytoskeletal proteins. They are hollow cylinders usually formed by 13 parallel protofilaments (PFs) covering the cylindrical walls of MTs. Each PF represents a series of proteins called tubulin dimers, with the length of $l = 8$nm [1-3]. They are electric dipoles whose longitudinal component of the electric dipole moment is $p = 337$Debye [4]. A segment of two neighbouring PFs is shown in Fig. 1. One can see that there is a longitudinal shift of $8l/13$ between neighbouring PFs [5,6].

A main purpose of this work is to introduce a new model. A crucial equation is solved both analytically and numerically. We show that a kink soliton is responsible for energy transfer along MT. To calculate total energy a useful theorem is introduced, which simplifies the calculation. On the basis of the results of this model we estimate wave velocity and wave width. Finally, to test the model, we calculate intrinsic electric field strength in two ways. Perfect agreement between the results suggests that the model is correct.

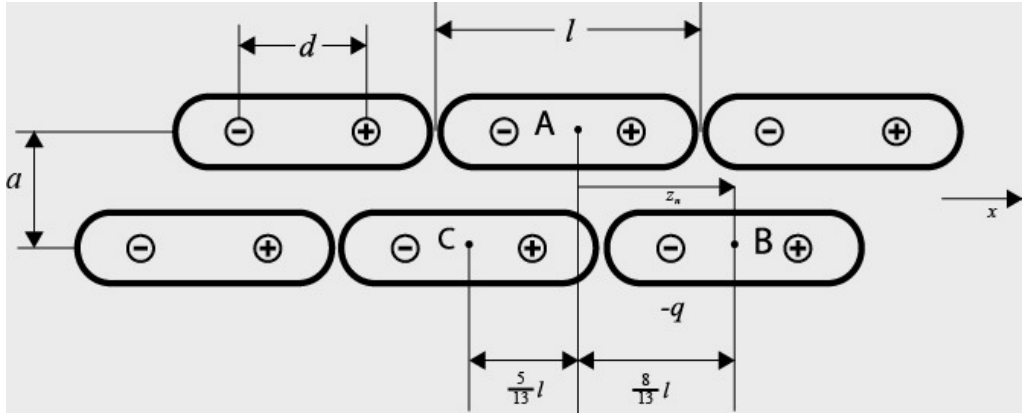

**Fig. 1.** A segment of two neighbouring protofilaments

## 2. Model

The starting point of the present modelling is the fact that the bonds between dimers within the same PF are significantly stronger than the soft bonds between neighbouring PFs [7,8]. This implies that the longitudinal displacements of pertaining dimers in a single PF should cause the longitudinal wave propagating along PF. The averaged impact of soft bonds with collateral PFs is taken to be described by the nonlinear double-well potential.

The present model assumes only one degree of freedom per dimer. This is $z_n$, a longitudinal displacement of a dimer at a position n.

The Hamiltonian for one PF is represented as

$$H = \sum_{n}\left[\frac{m}{2}\dot{z}_n^{\,2} + \frac{k}{2}(z_{n+1} - z_n)^2 + V(z_n)\right], \tag{1}$$

where dot means the first derivative with respect to time, $m$ is a mass of the dimer and $k$ is a harmonic constant describing the nearest–neighbour interaction between the dimers belonging to the same PF [9]. The first term represents a kinetic energy of the dimer, the second one, which we call harmonic energy, is a potential energy of the chemical interaction between the neighbouring dimers belonging to the same PF and the last term is the combined potential

$$V(z_n) = -Cz_n - \frac{1}{2}Az_n^2 + \frac{1}{4}Bz_n^4, \qquad C = qE, \tag{2}$$

where $E$ is the magnitude of the intrinsic electric field and $q$ represents the excess charge within the dipole. It is assumed that $q > 0$ and $E > 0$. One can recognize an energy of the dimer in the intrinsic electric field $E$ at the site $n$ and the well known double-well potential with positive parameters $A$ and $B$ that should be estimated [9].

The Hamiltonian given by Eqs. (1) and (2) is rather common in physics [9,10]. The first attempt to use it in nonlinear dynamics of MTs was done almost 20 years ago [9]. To be more precise, the Hamiltonian in Ref. [9] would be obtained from Eqs. (1) and (2) if $z_n$ were replaced by $u_n$. Hence, we refer to these two models as u-model and z-model. However, the meanings of $u_n$ in Ref. [9] and $z_n$ in the present paper are completely different. The u-model assumes an angular degree of freedom, while the coordinate $u_n$ is a projection of the top of the dimer on the direction of PF. On the other hand, the coordinate $z_n$ is a real displacement of the dimer along x axis. This will be further elaborated later.

Using generalized coordinates $z_n$ and $m\dot{z}_n$ and assuming a continuum approximation $z_n(t) \to z(x,t)$, we straightforwardly obtain the following nonlinear dynamical equation of motion

$$m\frac{\partial^2 z}{\partial t^2} - kl^2\frac{\partial^2 z}{\partial x^2} - qE - Az + Bz^3 + \gamma\frac{\partial z}{\partial t} = 0. \tag{3}$$

The last term represents a viscosity force with $\gamma$ being a viscosity coefficient [9].

It is well known that, for a given wave equation, a travelling wave $z(\xi)$ is a solution which depends upon $x$ and $t$ only through a unified variable $\xi \equiv \kappa x - \omega t$, where $\kappa$ and $\omega$ are constants. This allows us to obtain the final dimensionless ordinary differential equation

$$\alpha\psi'' - \rho\psi' - \psi + \psi^3 - \sigma = 0, \tag{4}$$

where $\psi' \equiv d\psi/d\xi$ and

$$\alpha = \frac{m\omega^2 - kl^2\kappa^2}{A}, \qquad z = \sqrt{\frac{A}{B}}\psi, \tag{5}$$

$$\rho = \frac{\gamma\omega}{A}, \qquad \sigma = \frac{qE}{A\sqrt{\frac{A}{B}}}. \tag{6}$$

The potential $V(z_n)$, in the continuum approximation, becomes

$$V(\psi) = \frac{A^2}{B}\left(-\sigma\psi - \frac{1}{2}\psi^2 + \frac{1}{4}\psi^4\right) \equiv \frac{A^2}{B}f(\psi). \tag{7}$$

The function $f(\psi)$ is shown in Fig. 2 for three different values of the parameter $\sigma$. The figure suggests that an increase in $\sigma$, i.e. in the value of $qE$, increases stability of MTs by lowering the right minimum. One can easily show that the values of $\psi$ for which $f(\psi)$ has right minimum, maximum and left minimum are respectively

$$\psi_R = \frac{2}{\sqrt{3}}\cos F, \quad \psi_M = \frac{1}{\sqrt{3}}\left(\sqrt{3}\sin F - \cos F\right), \quad \psi_L = -\frac{1}{\sqrt{3}}\left(\cos F + \sqrt{3}\sin F\right) \tag{8}$$

where

$$F = \frac{1}{3}\arccos\left(\frac{\sigma}{\sigma_0}\right), \qquad \sigma_0 = \frac{2}{3\sqrt{3}}. \tag{9}$$

Equation (4) is well known [9,10]. In what follow we solve it using an elegant mathematical method. Also, we will discus its new solution relevant for biological processes.

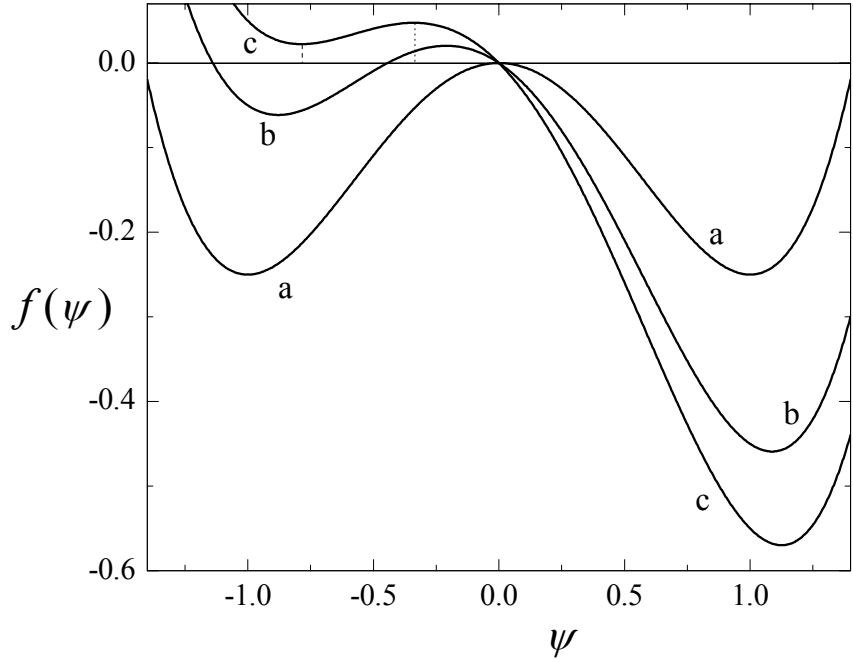

**Fig. 2.** The function $f(\psi)$ for: (a) $\sigma = 0$, (b) $\sigma = 0.2$ and (c) $\sigma = 0.35$.

## 3. Solution of Eq. (4)

A standard procedure for solving Eq. (4) is rather tedious [9,10]. A more elegant and relatively new method is the modified extended tanh-function method [11-15]. According to this procedure we look for a possible solution in the form

$$\psi = a_0 + \sum_{i=1}^{N_0}\left(a_i\Phi^i + b_i\Phi^{-i}\right), \tag{10}$$

where the function $\Phi = \Phi(\xi)$ is a solution of the well known Riccati equation

$$\Phi' = b + \Phi^2 \tag{11}$$

and $\Phi'$ is the first derivative [11,12]. The parameters $a_0$, $a_i$, $b_i$ and $b$ are real constants that should be determined as well as the integer $N_0$. The possible solutions of Eq. (11) depend on the sign of the parameter $b$. In our case $b < 0$ and $N_0 = 1$, which will be shown in a more elaborated version of the paper. This means that the function $\Phi = \Phi(\xi)$ is [11-15]

$$\Phi = -\sqrt{-b}\tanh\left(\sqrt{-b}\,\xi\right). \tag{12}$$

As $\Phi^{-1}$ diverges for small $\xi$ we look for the biophysically tractable solutions for which $a_1 \neq 0$ and $b_1 = 0$ in Eq. (10). All this brings about the following system of equations for the unknown parameters $b$, $a_0$, $a_1$ and $\alpha$

$$\left.\begin{aligned} -2a_0 + 8a_0^3 + \sigma &= 0, \\ -1 + 3a_0^2 + 2\alpha b &= 0, \\ \rho &= 3a_0 a_1, \\ 2\alpha &= -a_1^2. \end{aligned}\right\} \tag{13}$$

Notice that $a_0 a_1 > 0$ and $\alpha < 0$. Negative $\alpha$ means that the harmonic energy in the Hamiltonian (1) is larger than the kinetic energy. This may indicate a small velocity of the wave and/or a big $k$, which can be seen from Eq. (5). As the neighbouring dimers are not connected by strong chemical bonds, we expect very small wave velocity. Notice, also, that $b < 0$ holds for the inequality $a_0^2 < 1/3$, which is equivalent to $\sigma < \sigma_0$ in Eq. (9).

Three real values of $a_0$ satisfy the first of Eq. (13). They describe the system in the states corresponding to the right and left minimum and the maximum (See Fig. 2). This will be derived and important details will be explained in a more elaborated version of the paper. It suffices now to mention that the function $\psi_1(\xi)$, describing the system in the right minimum, is

$$\psi_1(\xi) = a_0 - b_0 \tanh\left(\frac{3a_0 b_0}{\rho}\xi\right), \tag{14}$$

where

$$a_0 \equiv a_{01} = \frac{1}{2\sqrt{3}}\left(\cos F + \sqrt{3}\sin F\right), \qquad b_0 = \sqrt{1 - 3a_0^2}. \tag{15}$$

We can see that $\psi_1(\xi)$ is an antikink soliton depending on both $\rho$ and $\sigma$, as $a_{01}$ is a function of $\sigma$. Its asymptotic values are $\psi_1(-\infty) = \psi_R$ and $\psi_1(+\infty) = \psi_M$. Notice that the jump of the function $\psi_1(\xi)$ depends on $\sigma$ only, while the solitonic width, i.e. its slope, depends on both $\rho$ and $\sigma$. It is obvious that the solitonic width is proportional to viscosity.

The same solution of Eq. (4), i.e. Eqs. (14) and (15), can be obtained using Jacoby elliptic functions. This method will be exploited and explained in a separate publication.

It is very important to keep in mind that the solution given by Eqs. (14) and (15) is different from what was offered before [9,10]. To study two ground states of a proton [10] the asymptotic values $\psi_2(-\infty) = \psi_R$ and $\psi_2(+\infty) = \psi_L$ were assumed. Of course, both states $\psi_R$ and $\psi_L$ are stable. However, the fact that the state $\psi_M$ is unstable has an important biophysical meaning. Suppose that a dimer obtained a portion of energy released by guanosine triphosphate (GTP), which results in its displacement. This displacement affects the surrounding dimers causing the wave creation. If the dimer moves from $\psi_R$ to $\psi_L$ it needs one more portion of energy to escape from $\psi_L$. We do expect that the rational Nature offers only one energy supply and that the system moves from the bottom of Fig. 2 to a value that is close but bellow the maximum value. In other words, the transition from $\psi_R$ to any state close to $\psi_M$ is enough for the soliton creation and allows a spontaneous return to the ground state $\psi_R$.

The function $\psi_1(\xi)$ is shown in Fig. 3 together with the results of numerical solutions of Eq. (4). Perfect agreement between the respective curves certainly validates our analytical approach. For the numerical solution the standard shooting method with the 4$^{th}$ order Runge-Kutta integrator was used. Particular solutions are determined by their asymptotic behaviour and are centred at $\xi = 0$. The numerical step was $10^{-5}$.

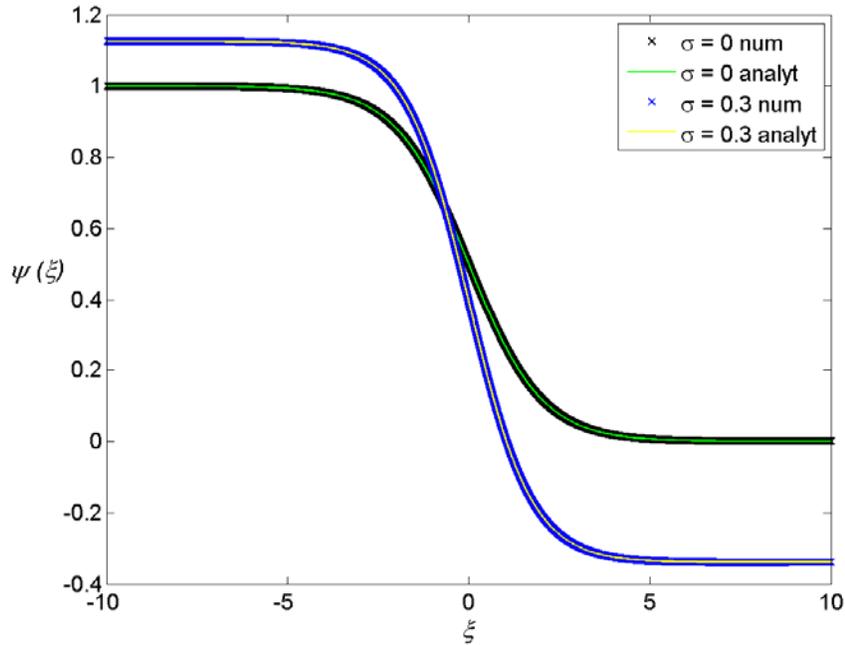

**Fig. 3.** The function $\psi_1(\xi)$ for $\rho = 1.5$ and: (a) $\sigma = 0$ and (b) $\sigma = 0.3$.

Also, Eq. (4) was numerically solved for some values of the parameter $\alpha$ for which our analytical procedure does not give solutions. This is shown in Fig. 4. One can see that for small values of $|\alpha|$ the transition between the initial and the final position of the dimer is fast and smooth. Otherwise, for big $|\alpha|$ the dimer experiences oscillation before it stabilizes in the final value. These results indicate that there is an optimal parameter $\alpha$ for a given process to occur.

In what follows we will estimate kink's velocity and some other parameters. For this purpose we should calculate the total system's energy. This can be simplified using the following theorem.

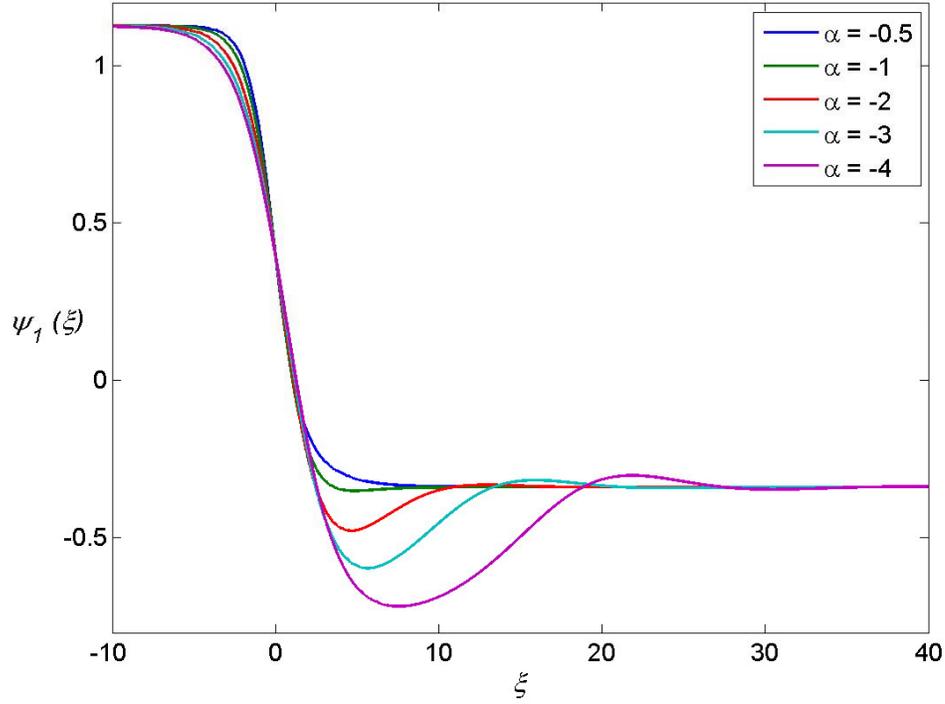

**Fig. 4.** The function $\psi_1(\xi)$ for $\rho = 1.5$ and $\sigma = 0.3$ for different values of $\alpha$

### 4. Theorem

Let:

1. $$H = \sum_n \left[ \frac{m}{2} \dot{q}_n^2 + \frac{k}{2}(q_{n+1} - q_n)^2 + f(q_n) \right] \quad (16)$$

represents Hamiltonian of one dimensional system where dot means the first derivative with respect to time, the integer $n$ determines the position of the considered particle and $m$ and $k$ are constants,

2. a continuum approximation is valid for the system,
3. the function $f$ is an arbitrary differentiable function,
4. $q(x,t) = q(\xi)$, $\xi = \kappa x - \omega t$, $\quad (17)$

and

    5. the integral $\int_{-\infty}^{\infty} f(q)dx$ converges.

Let the total energy be $\mathcal{E} = \mathcal{E}_1 + \mathcal{E}_2 + \mathcal{E}_3$, where the energies $\mathcal{E}_1, \mathcal{E}_2, \mathcal{E}_3$ respectively correspond to the three terms in Eq. (16). Then

$$\mathcal{E} = 2\mathcal{E}_2, \quad \text{i.e.} \quad \mathcal{E}_3 = \mathcal{E}_2 - \mathcal{E}_1. \tag{18}$$

## 5. Proof

We use the Hamilton equation $\dot{p}_n = m\ddot{q}_n = -\partial H/\partial q_n$, a series expansion of $q_{n\pm 1}$, a continuum approximation $q_n(t) \to q(x,t)$ and Eq. (17), which bring about the following equation

$$\left(m\omega^2 - kl^2\kappa^2\right)\frac{d^2q}{d\xi^2} + \frac{df}{dq} = 0. \tag{19}$$

The total energy, in continuum approximation, can be calculated as

$$\mathcal{E} = \frac{1}{l}\int_{-\infty}^{\infty} \varepsilon \, dx \tag{20}$$

where

$$\varepsilon = \frac{m}{2}\left(\frac{\partial q}{\partial t}\right)^2 + \frac{kl^2}{2}\left(\frac{\partial q}{\partial x}\right)^2 + f(q) \equiv \varepsilon_1 + \varepsilon_2 + \varepsilon_3. \tag{21}$$

Using Eq. (17) we easily obtain

$$\varepsilon = \left(\frac{m\omega^2}{2} + \frac{kl^2\kappa^2}{2}\right)\left(\frac{dq}{d\xi}\right)^2 + f(q) \tag{22}$$

as well as

$$\frac{d\varepsilon}{d\xi} = \left[\left(m\omega^2 + kl^2\kappa^2\right)\frac{d^2q}{d\xi^2} + \frac{df}{dq}\right]\frac{dq}{d\xi}. \tag{23}$$

Eqs. (19) and (23) give

$$\frac{d\varepsilon}{d\xi} = 2kl^2\kappa^2 \frac{d^2q}{d\xi^2}\frac{dq}{d\xi} = kl^2\kappa^2 \frac{d}{d\xi}\left(\frac{dq}{d\xi}\right)^2. \tag{24}$$

If we set a constant of integration to be zero, for the integral (20) to converge, as required by the point five of the theorem, we obtain

$$\varepsilon = kl^2\kappa^2 \left(\frac{dq}{d\xi}\right)^2 = 2\varepsilon_2, \tag{25}$$

which proves the theorem.

## 6. Estimations

*6.1. Electric field strength*

We want to calculate a projection of electric field on the x-axis in the point A in Fig. 1, which is denoted as $E$ in Eq. (2). One way to calculate this internal electric field is straightforward but tedious. We can calculate the components coming from all the excess charges in MT. For example, electric field from the denoted charge $-q$ in Fig. 1 is

$\frac{q}{4\pi\varepsilon_0\varepsilon_r}\left(\frac{8l}{13}-\frac{d}{2}\right)\left[\left(\frac{8l}{13}-\frac{d}{2}\right)^2 + a^2\right]^{-\frac{3}{2}}$. Non-negligible contributions are from PF including the dimer with the point A and from the two neighbouring pairs of PFs. For $\varepsilon_r = 8.41$, $q = 1.75e$, $d = 4.04$nm [4] and for $a = 0.60l$, which comes from the geometry of MT as the inner and the outer diameters are 15nm and 25nm [5], we finally obtain

$$E = 1.7 \times 10^7 \text{ N/C}. \tag{26}$$

Another way to estimate $E$ is related to the meaning of $z_n$. This estimate is extremely important, representing one of the tests of the model introduced in this paper. It was pointed out that $z_n$ was a longitudinal displacement of the dimer. This should be understood as the displacement of the dimer with respect to its neighbour from the neighbouring PF. (See Fig. 1). This, practically, means that the two minima in Fig. 2 correspond to the dimers shifted by $8l/13$ and $-5l/13$ (See Fig. 1), yielding the following system of equations

$$\sqrt{\frac{A}{B}}\psi_R = \frac{8l}{13}, \qquad \sqrt{\frac{A}{B}}\psi_L = \frac{-5l}{13}. \tag{27}$$

A solution of this system is $\sigma = 0.91\sigma_0$ and $\sqrt{A/B} = 4.31\text{nm}$. Unfortunately, this is not enough to calculate $E$ as the value of the parameter $A$ should be known, which can be seen from Eq. (6). Notice rather large $\sigma$, which brings about the negligible left minimum (line c in Fig. 2).

A next step is to calculate $A$. It is well known that MTs serve as a "road network" for motor proteins (kinesin and dynein) dragging different "cargos" such as vesicles and mitochondria to different sub-cellular locations. The anti-kink soliton from above should be understood as a signal for the protein to start moving along PF. The required energy for the soliton creation comes from hydrolysis of GTP. When the system obtains the portion of energy it moves from the bottom in Fig. 2 to a higher energy level. If we suppose that the energy released by GTP, which is $0.25\text{eV}$, matches the difference between the minimum and maximum, i.e.

$$|V(\psi_R)| + V(\psi_M) = 0.25\text{eV}, \tag{28}$$

we obtain the value

$$A = 3.1 \times 10^{-3} \text{ N/m}. \tag{29}$$

This brings about exactly the same value for $E$ as already calculated, i.e. the value given by Eq. (26), which indirectly validates our model.

A careful reader may notice that the GTP energy affects the total energy rather than the potential energy $V$ only. Maximum of $V$ corresponds to the zero value of kinetic energy, but the harmonic term, being essentially the potential energy, deserves a comment. In our approach the first two terms in Eq. (3) yield the first term in Eq. (4) as both of them are proportional to $\partial^2 z/\partial \xi^2$. This practically means that the harmonic term should be treated as the kinetic energy and that our estimation (29) is correct.

*6.2. Kink velocity and kink width*

From Eq. (14) we can recognize the solitonic width $\Lambda \equiv Nl$, with $N$ being a number of dimers, as [16]

$$\frac{2\pi}{Nl} = \frac{3a_0\kappa}{\rho}b_0, \tag{30}$$

which, together with the first of Eqs. (6), gives

$$v \equiv \frac{\omega}{\kappa} = \frac{3a_0 l}{2\pi\gamma} b_0 AN. \tag{31}$$

Also, according to the expressions for $\alpha$ and $\rho$ in Eqs. (5) and (6) and the last two of Eq. (13) we easily obtain

$$k = \frac{v^2}{l^2}(m + \frac{\gamma^2}{18Aa_0^2}) \equiv \frac{v^2}{l^2}(m + M) \approx \frac{v^2}{l^2}M. \qquad (32)$$

For the viscosity coefficient $\gamma$ we use $\gamma = 5.3 \times 10^{-11}$ kg/s [9], which is still a disputable value.

Using the theorem proved above we can derive the following expression for the total energy

$$\mathcal{E} = \frac{4kl\kappa a_0}{\rho}\frac{A}{B}b_0^3. \qquad (33)$$

Equations (31)-(33) and (6) yield the value

$$\mathcal{E} = \frac{1}{3\pi}\frac{A}{B}b_0^4 AN = 4.70AN, \qquad (34)$$

where the units $\mathcal{E}_u = \text{eV}$ and $A_u = \text{N/m}$ are assumed. To derive Eq. (33) we incorporated the following constant value to the potential $V$, initially given by Eq. (2), which is convenient for integration:

$$\Delta V = qE\sqrt{\frac{A}{B}}a_0 + \frac{A^2}{2B}(a_0^2 + b_0^2) - \frac{A^2}{4B}(a_0^4 + 6a_0^2 b_0^2 + b_0^4). \qquad (35)$$

This ensures the point five in the theorem from above to be satisfied. The minimum of $V$ is $V(\psi_R) = -0.125\text{eV}$. The GTP energy is delivered to a single dimer and is transferred to $N$ of them due the interactions between the dimers. Therefore, we can write

$$4.70AN + 0.125 = 0.25. \qquad (36)$$

The obtained value for $AN$, together with Eqs. (29) and (31), gives

$$v = 0.54\,\text{m/s} \qquad (37)$$

and

$$N = 8.6, \qquad (38)$$

which means that about nine dimers are covered by the kink.

## 7. Concluding remarks

According to our model the kink soliton is responsible for the energy transfer along the MT. Such wave can be utilized by a cell for some important mechanisms underlying its functional dynamics. As MTs serve as a "road network" for motor proteins it is very likely that the cell's compartment which needs the specific cargo will launch the soliton of above kind sending it as a signal along the closest MT. The soliton will activate proper motors being close to MT along which the soliton propagates.

Our model allows us to make some estimates. The small value for $v$ was anticipated when we discussed the negative sign of $\alpha$. One should keep in mind that the forces responsible to the soliton creation are not strong covalent bonds, which would provide big $v$. According to the model the kink covers about nine dimers. This result is reasonable as very small $N$ would be in contradiction with the continuum approximation. To test the validity of this approximation a discrete counterpart of Eq. (3) should be solved numerically and the results should be compared. This research is in progress and will be submitted soon.

**Acknowledgements**

S.Z. thanks to mathematician Professor Stana Nikčević for her help and to Dr. Zoran Ivić for useful discussions. S.Z. also thanks to Dejan Milutinović who plotted Fig. 1. This research was supported by funds from Serbian Ministry of Education and Sciences, grants III45010 and OI171009.